\documentclass[preprints,article,accept,pdftex,moreauthors]{Definitions/mdpi} 

\usepackage{xcolor}

\firstpage{1} 
\pubvolume{1}
\issuenum{1}
\articlenumber{0}
\pubyear{2022}
\copyrightyear{2022}

\datereceived{} 
\dateaccepted{} 
\datepublished{} 

\usepackage[flushleft]{threeparttable}
\usepackage{graphicx}
\usepackage{dcolumn}
\usepackage{bm}
\usepackage{amsfonts}
\usepackage{amsmath}
\usepackage{amssymb}
\usepackage[utf8]{inputenc}
\usepackage[T1]{fontenc}

\hreflink{https://doi.org/} 

\pdfoutput=1

\Title{Numerical Simulations of Dark Matter Admixed Neutron Star Binaries}

\TitleCitation{Title}

\Author{Mattia Emma $^{1}$, Federico Schianchi$^1$\orcidD{}, Francesco Pannarale$^{2,3}$\orcidC{}, Violetta Sagun$^{4}$\orcidB{}, Tim Dietrich$^{1,5}$\orcidA{}}

\AuthorNames{Mattia Emma, Federico Schianchi, Francesco Pannarale, Violetta Sagun, Tim Dietrich}

\AuthorCitation{Emma, M.; Schainchi, F.; Pannarale, F.; Sagun, V.; Dietrich, T.}

\address{$^{1}$ \quad Institut f\"{u}r Physik und Astronomie, Universit\"{a}t Potsdam, Haus 28, 14476 Potsdam, Germany; \\
$^{2}$ \quad Dipartimento di Fisica, Universit\`{a} di Roma ``Sapienza'', Piazzale A. Moro 5, I-00185, Roma, Italy; \\
$^{3}$ \quad INFN Sezione di Roma, Piazzale A. Moro 5, I-00185, Roma, Italy; \\ 
$^{4}$ \quad CFisUC, University of Coimbra, Rua Larga 3004-516, Coimbra, Portugal; \\
$^{5}$ \quad Max Planck Institute for Gravitational Physics (Albert Einstein Institute), Am M\"{u}hlenberg 1, 14476 Potsdam, Germany; \\}

\corres{Correspondence: mattia.emma@uni-potsdam.de}

\abstract{
Multi-messenger observations of compact binary mergers provide a new way to constrain the nature of dark matter that may accumulate in and around neutron stars. In this article, we extend the infrastructure of our numerical-relativity code BAM to enable the simulation of neutron stars that contain an additional mirror dark matter component. We perform single star tests to verify our code and the first binary neutron star simulations of this kind. We find that the presence of dark matter reduces the lifetime of the merger remnant and favors a prompt collapse to a black hole. Furthermore, we find differences in the merger time for systems with the same total mass and mass ratio, but different amounts of dark matter. Finally, we find that electromagnetic signals produced by the merger of binary neutron stars admixed with dark matter are very unlikely to be as bright as their dark matter free counterparts. Given the increasing sensitivity of multi-messenger facilities, our analysis gives a new perspective on how to probe the presence of dark matter.}

\keyword{numerical relativity, dark matter, neutron stars, equation of state, gravitational-wave astronomy, multi-messenger astrophysics.} 

\begin{document}

\section{Introduction}
\label{sec:Intr}

Neutron stars, which are among the densest and most compact astrophysical objects, enable studies of the properties of strongly interacting matter under extreme conditions. In this regard, the first detection of the gravitational-wave (GW) signal emitted by a binary neutron star coalescence~\cite{LIGOScientific:2017vwq} and of its associated electromagnetic counterparts~\cite{LIGOScientific:2017ync, LIGOScientific:2017zic} mark a new era in GW and multi-messenger astronomy, disciplines that will further improve our understanding of matter at supranuclear densities. 

Extracting the source properties of the binary from the observational data requires accurate theoretical models of the last stages of the binary coalescence. Due to the complexity of the equations governing general relativity and relativistic fluids, these can only be achieved with the aid of numerical-relativity simulations; cf.~\cite{Baumgarte:2010,Rezzolla:2013,Shibata:2016} for textbook discussions and further references. While these simulations have undergone significant improvements over the last decades, the impact of dark matter on the binary neutron star dynamics has not been investigated in full detail yet, but cf.~among others \cite{Bauswein:2020kor,Miao:2022rqj,Karkevandi:2021ygv} for first studies in this direction.

Numerous studies suggest that dark matter can cluster around and inside compact objects such as neutron stars, e.g., \cite{2006PhRvD..74f3003N,Bertone:2007ae,2010PhRvD..81l3521D,2010PhRvD..82f3531K,2011PhRvD..84j7301L,2012PhLB..711....6P,2013arXiv1308.3222K,2015PhRvD..92f3526K,2018PhRvD..97l3007E,2020MNRAS.495.4893D,Sagun:20224z,2022arXiv220307984B}. As a result, dark matter may increase the total gravitational mass of a neutron star due to an extended halo formation or, on the contrary, reduce it, due to the formation of a dense core, e.g., \cite{Ellis:2018bkr, Ivanytskyi:2019wxd,2019JCAP...07..012N,Karkevandi:2021ygv}. Which scenario occurs depends on the dark matter properties, such as the dark matter particle mass, the self-interacting coupling constant and its relative abundance inside the star. In the case of binary neutron star systems, both scenarios can impact the measured tidal deformability~\cite{Ciancarella:2020msu} and the merger dynamics. Hence, an analysis of the emitted GW signal provides a unique opportunity to probe dark matter in neutron star mergers.  

Similarly, matter outflows, driven by tidal disruption, shock heating, and disk wind, depend not only on the total mass, mass ratio, and spin of the two neutron stars, but also on the equation of state (EoS) describing neutron star matter, e.g., \cite{Hotokezaka:2012ze,Bauswein:2013yna,Dietrich:2016fpt,Radice:2017lry,Fujibayashi:2017xsz,Nedora:2020qtd,Kruger:2020gig}, which may be significantly altered due to the presence of dark matter in the star interior. Hence, also the observation of electromagnetic counterparts, triggered by the outflowing material, might provide evidence for the existence or absence of dark matter during the binary neutron star coalescence. 

Although speculative, one of the first candidates of a potential neutron star admixed with dark matter is the secondary component of GW190814 \cite{LIGOScientific:2020zkf}. An interpretation of the observed compact object with a mass of $\sim2.6 ~M_{\odot}$ as a canonical neutron star would raise a lot of difficulties for existing models of baryon matter, e.g., \cite{Fattoyev:2020cws,Tsokaros:2020hli,Zhang:2020zsc,Tan:2020ics}, hence, it could be a dark matter admixed neutron star. The accumulation of dark matter \cite{Karkevandi:2021ygv,DiGiovanni:2021ejn,2021PhRvD.104f3028D} onto the compact star provides a simple explanation that keeps the baryonic matter models in agreement with GW190814. However, also the interpretation of GW190814 as a binary black hole merger is certainly possible and likely, e.g.,~\cite{Essick:2020ghc,Tews:2020ylw}.

Up to our knowledge, Refs.~\cite{Ellis:2017jgp,2018JHEP...11..096K,Bezares:2019jcb,Bauswein:2020kor} report the first analyses of the effects of dark matter on the GW signal produced by the merger of binary neutron star with dark matter and, particularly, on the postmerger part of the waveform. In this work we go further by performing, up to our knowledge, the first two-fluid 3D simulations of coalescencing binary neutron star systems admixed with dark matter. As a first step, we model dark matter as mirror dark matter. This specific dark matter candidate mirrors the baryon matter to a parallel hidden sector, which implies the same particle physics as the observable world and couples to the latter through gravity~\cite{2004IJMPA..19.3775B}.\footnote{
We plan to investigate more elaborate dark matter models in future works, but for simplicity consider mirror dark matter as a starting point.}

The baryonic matter component employed in this work is described by the piecewise-polytropic fit ~\cite{Read:2008iy} of the SLy EoS~\cite{Douchin:2001sv} that reproduces nuclear matter ground state properties, and fulfills tidal deformability constraints \cite{LIGOScientific:2017vwq, LIGOScientific:2018hze} as well as the $2~M_{\odot}$ limit \cite{Antoniadis:2013pzd, NANOGrav:2019jur}.

This article is structured as follows.  First, in Sec.~\ref{sec:Meth}, we discuss the numerical-relativity code BAM and the implementation of mirror dark matter within it. In Sec.~\ref{sec:Sim} we construct a single stable mirror dark matter admixed neutron star and study its stability during dynamical simulations. In Sec.~\ref{sec:Bin} the construction of initial configurations and results for simulations of binary systems are presented.  Our conclusions follow in Sec.~\ref{sec:Conc}.

\section{Materials and Methods}
\label{sec:Meth}

\subsection{BAM code }
\label{sec:BAM}

In this work, we employ the numerical-relativity code BAM~\cite{Bruegmann:2006ulg,Thierfelder:2011yi,Dietrich:2015iva,Dietrich:2018bvi}, which allows us to solve the 3+1-decomposed Einstein Equations coupled with the equations describing general-relativistic hydrodynamics (GRHD). 
BAM is based on the method of lines, it employs finite differencing to discretize the spacetime, and it uses high-resolution shock capturing methods to ensure a robust evolution of relativistic fluids~\cite{Thierfelder:2011yi, Font:2000pp}. 
The time evolution is performed with an explicit time integrator, e.g., Runge-Kutta. BAM employs an adaptive mesh technique based on a set of Cartesian boxes, while the time integration uses the Berger-Oliger~\cite{1989JCoPh..82...64B, 1984JCoPh..53..484B} or Berger-Collela algorithm~\cite{Dietrich:2015iva}. 

BAM supports the BSSN~\cite{Baumgarte:1998te, Shibata:1995we} and Z4c~\cite{Bernuzzi:2009ex, Hilditch:2012fp, Weyhausen:2011cg} formulations of the Einstein Equations. Here, we use the latter because of its constraint damping properties that reduce the initial constraint violations that are present in our binary neutron star setups. In all our simulations we use the moving puncture gauge~\cite{vanMeter:2006vi}. 

In addition to the simulation of relativistic fluids and the Einstein Equations, BAM has already been used to simulate potential dark matter components. Previous works included the simulation of boson and axion stars~\cite{Dietrich:2018bvi,Clough:2018exo,Dietrich:2018jov}, and binary neutron star simulations accounting for cooling through nucleon-nucleon-axion bremsstrahlung~\cite{Dietrich:2019shr}. 
In this article, we follow a different approach and consider mirror dark matter as the possible explanation of the dark matter that is present in our Universe~\cite{Foot:2014mia}. Mirror dark matter constitutes only a first natural step to extend our code infrastructure to allow for the simulation of more complex dark matter models that we plan to investigate in the future. 

\subsection{Inclusion of mirror dark matter}          
We use a two-fluid approach for the inclusion of mirror dark matter. For this purpose, we double the hydrodynamic variables, i.e., the primitive variables \{$p, \rho, \epsilon, \Vec{v}$\}, the pressure, density, specific internal energy and velocity, and the conserved variables \{$D, \Vec{S},\tau$\}, the conserved rest-mass density, momentum density, and internal energy density of the system from an Eulerian observer’s view.
Each parameter of `visible` matter obtains its mirror (`dark') counterpart. 

Given the assumption that dark matter couples to normal, baryonic matter only through gravity, we are able to evolve both fluid components independently from one another.
This reflects in the ansatz that the total stress-energy tensor is the sum of the two components, so that the Einstein Equations read
\begin{linenomath}
\begin{equation}
R_{\mu \nu } - \frac{1}{2} R g_{\mu \nu} = 8 \pi T_{\mu \nu}= 8 \pi \textcolor{blue}{T_{\mu \nu}^{\rm b}}+ 8 \pi  \textcolor{red}{T_{\mu \nu }^{\rm dm}},
\label{tmunu}
\end{equation} 
\end{linenomath}
where \textcolor{blue}{b} stands for the normal baryonic matter and \textcolor{red}{dm} for the dark matter. In the following, we will use the introduced labels and color scheme to distinguish the two fluids.  
Given that the normal and mirror dark matter interact only through gravity, the fluid degrees of freedom decouple and the ADM (Arnowitt-Deser-Misner) variables can simply be considered as the sum of the single parameters, e.g., $\rho_{ADM}=\textcolor{blue}{\rho_{\rm b}}+\textcolor{red}{\rho_{\rm dm}}$.

This implies that we double the hydrodynamical evolution equations, which appear in the code in their 3+1 decomposed form: 
\begin{linenomath}
\begin{align}
   \partial_t \textcolor{blue}{ D_{\rm b}}-\beta^k\partial_k\textcolor{blue}{ D_{\rm b}}+\textcolor{blue}{D_k^{\rm b}}(\alpha \textcolor{blue}{D_{\rm b} v^k_{\rm b}})&=\alpha K \textcolor{blue}{D_{\rm b}}, 
   \\ 
   \partial_t \textcolor{red}{ D_{\rm dm}}-\beta^k\partial_k\textcolor{red}{ D_{\rm dm}}+\textcolor{red}{D_k^{\rm dm}}(\alpha \textcolor{red}{D_{\rm dm} v^k_{\rm dm}})&=\alpha K \textcolor{red}{D_{\rm dm}},\nonumber
    \end{align}
\end{linenomath}
\begin{linenomath}
{\small
\begin{align}
    \alpha^2[\textcolor{blue}{T^{0\mu}_{\rm b}}\partial_{\mu}(\ln\alpha)-\Gamma_{\mu \nu}^0\textcolor{blue}{T^{\mu \nu}_{\rm b}}]&= (\textcolor{blue}{\tau_{\rm b}}+\textcolor{blue}{p_{\rm b}}+ \textcolor{blue}{D_{\rm b}})(\alpha\textcolor{blue}{v^m_{\rm b}v^n_{\rm b}}K_{mn}-\textcolor{blue}{v^m_{\rm b}}\partial_m\alpha)+\alpha \textcolor{blue}{p_{\rm b}}K+\alpha K \textcolor{blue}{\tau_{\rm b}},
    \\
    \alpha^2[\textcolor{red}{T^{0\mu}_{\rm dm}}\partial_{\mu}(\ln\alpha)-\Gamma_{\mu \nu}^0\textcolor{red}{T^{\mu \nu}_{\rm dm}}]&= (\textcolor{red}{\tau_{\rm dm}}+\textcolor{red}{p_{\rm dm}}+ \textcolor{red}{D_{\rm dm}})(\alpha\textcolor{red}{v^m_{\rm dm}v^n_{\rm dm}}K_{mn}-\textcolor{red}{v^m_{\rm dm}}\partial_m\alpha)+\alpha \textcolor{red}{p_{\rm dm}}K+\alpha K\textcolor{red}{\tau_{\rm dm}},
    \nonumber
\end{align}
}%
\end{linenomath}

\begin{linenomath}
\begin{align}
\partial_t \textcolor{blue}{S^i_{\rm b}}-\mathcal{L}_{\Vec{\beta}}\textcolor{blue}{S^i_{\rm b}}+\textcolor{blue}{D_k^{\rm b}}[\alpha(\textcolor{blue}{S^i_{\rm b}v^k_{\rm b}}+\gamma^{ik}\textcolor{blue}{p_{\rm b}})]&=-(\textcolor{blue}{\tau_{\rm b}}+\textcolor{blue}{D_{\rm b}})\textcolor{blue}{D^i_{\rm b}}\alpha+\alpha K \textcolor{blue}{S^i_{\rm b}}, \\
\partial_t \textcolor{red}{S^i_{\rm dm}}-\mathcal{L}_{\Vec{\beta}}\textcolor{red}{S^i_{\rm dm}}+\textcolor{red}{D_k^{\rm dm}}[\alpha(\textcolor{red}{S^i_{\rm dm}v^k_{\rm dm}}+\gamma^{ik}\textcolor{red}{p_{\rm dm}})]&=-(\textcolor{red}{\tau_{\rm dm}}+\textcolor{red}{D_{\rm dm}})\textcolor{red}{D^i_{\rm dm}}\alpha+\alpha K \textcolor{red}{S^i_{\rm dm}}.\nonumber
\end{align}
\end{linenomath}

To ensure the stability of the simulation, we also double the artificial atmosphere that surrounds the neutron stars in normal GRHD simulations, but we use a different threshold value for the two densities, following the single fluid procedure outlined in~\cite{Thierfelder:2011yi} for each individual component. 

\section{Single Star Simulations}
\label{sec:Sim}

\subsection{Solving the TOV Equations}

To construct single star data, we have to adjust the TOV (Tolman–Oppenheimer–Volkoff) equations \cite{PhysRev.55.364,PhysRev.55.374} to produce a stable mirror dark matter admixed neutron star. 
As the TOV equations result from imposing the conservation of mass and energy $\nabla_{\nu}T^{\mu \nu }=0$, this becomes 
\begin{linenomath}
\begin{equation}
    \textcolor{blue}{\nabla_{\nu} T^{\mu \nu}_{\rm b}}+\textcolor{red}{\nabla_{\nu} T^{\mu \nu}_{\rm dm}}=0  ,
    \label{divt}
\end{equation}
\end{linenomath}
where, since the two components are only coupled through gravity, they are also individually conserved. Hence, the TOV equations become
\begin{linenomath}
\begin{equation}
  \textcolor{blue}{\frac{d\rho_{\rm b}}{dr}= -(\epsilon_{\rm b}+p_{\rm b})}\frac{d\phi}{dr}\textcolor{blue}{\frac{d\rho_{\rm b}}{dp_{\rm b}}}  ,
   \quad 
   \textcolor{red}{\frac{d\rho_{\rm dm}}{dr}= -(\epsilon_{\rm dm}+p_{\rm dm})}\frac{d\phi}{dr}\textcolor{red}{\frac{d\rho_{\rm dm}}{dp_{\rm dm}}}  ,
   \label{tov1}
\end{equation} 

\begin{equation}
   \textcolor{blue}{\frac{dm_{\rm b}}{dr}}= 4\pi r^2\textcolor{blue}{\epsilon_{\rm b}}  , 
    \quad  \textcolor{red}{\frac{dm_{\rm dm}}{dr}}= 4\pi r^2\textcolor{red}{\epsilon_{\rm dm}}  ,
    \label{tov2}
\end{equation} 

\begin{equation}
    \frac{d\phi}{dr}= \frac{M+4\pi r^3P}{r^2(1-\frac{2M}{r})} ,
\end{equation} 
\end{linenomath}
where $\frac{d\phi}{dr}$ ensures the gravitational coupling between the two fluids,  $P=\textcolor{blue}{p_{\rm b}}+\textcolor{red}{p_{\rm dm}}$ is the total pressure and $M$ the total mass with contributions from both species of matter. 
To obtain a unique solution of the described system of equations, the initial central densities (of the dark and visible matter) have to be known.

When solving the system of equations, we integrate as long as at least one of the two densities is non-zero. When one of the two densities vanishes, the density is set to the atmosphere value and kept constant throughout the rest of the integration, while the integration continues for the other component. This makes it possible to build and evolve models of neutron stars with an arbitrary amount of mirror dark matter.
\begin{figure}[htp!]
\begin{center}
\includegraphics[width=0.49\textwidth]{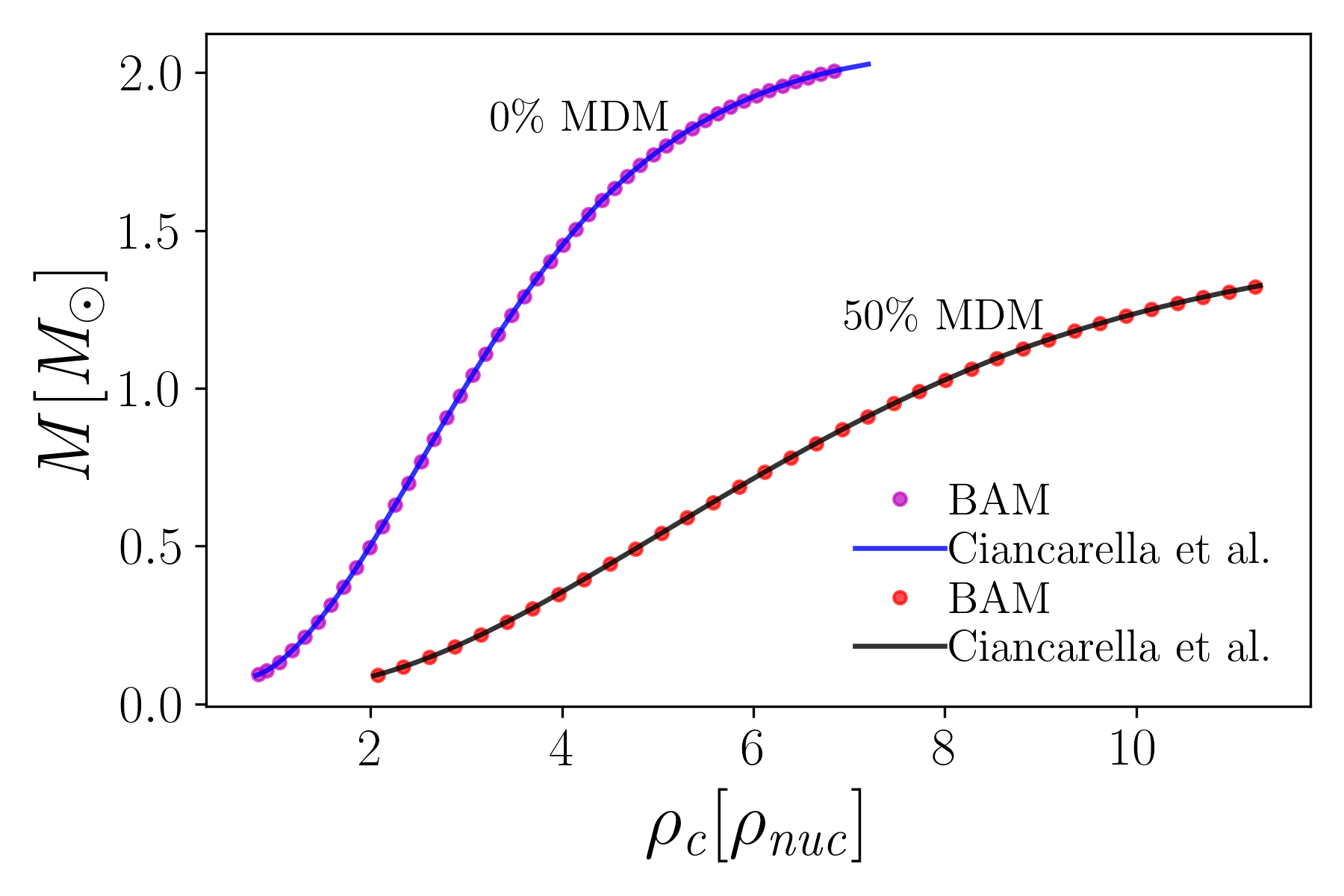}
\includegraphics[width=0.49\textwidth]{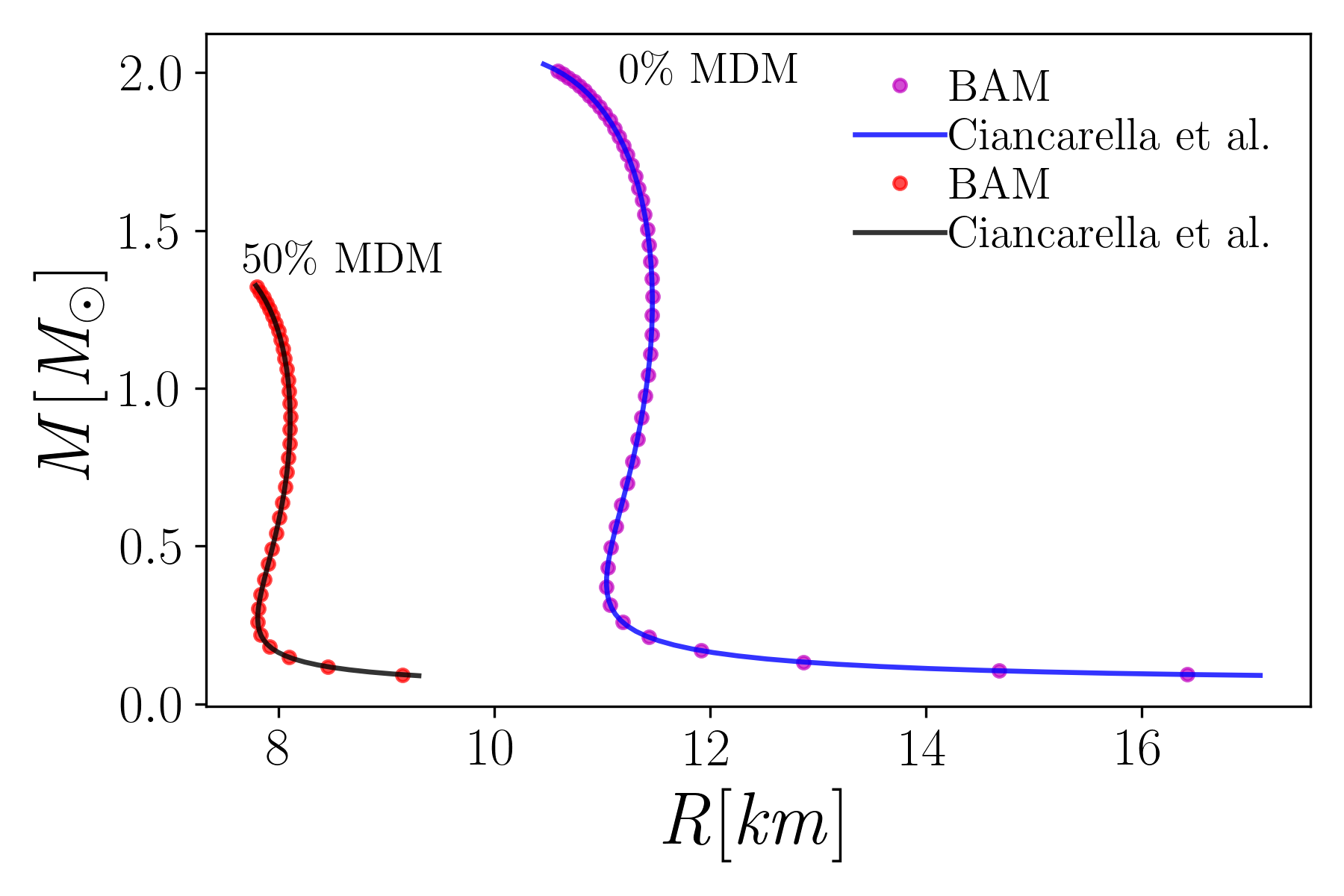}
\caption{ \textbf{Left}: Total mass values as a function of the total central density of the neutron star, for a pure baryonic star (blue line, magenta dots) and 50\% mirror dark matter admixed star (black line, red dots).  The density is expressed in units of the nuclear saturation denisty. The SLy EoS is used for both plots.
\textbf{Right}: Total mass as a function of the radius, for the same neutron star models.\\ The lines show data obtained as described in Ciancarella et al.~\cite{Ciancarella:2020msu} and the overplotted dots result from our updated BAM version. \label{fig:density_radius}}
\end{center}
\end{figure} 

\begin{figure}[htp!]
\begin{center}
\includegraphics[width=0.54 \textwidth]{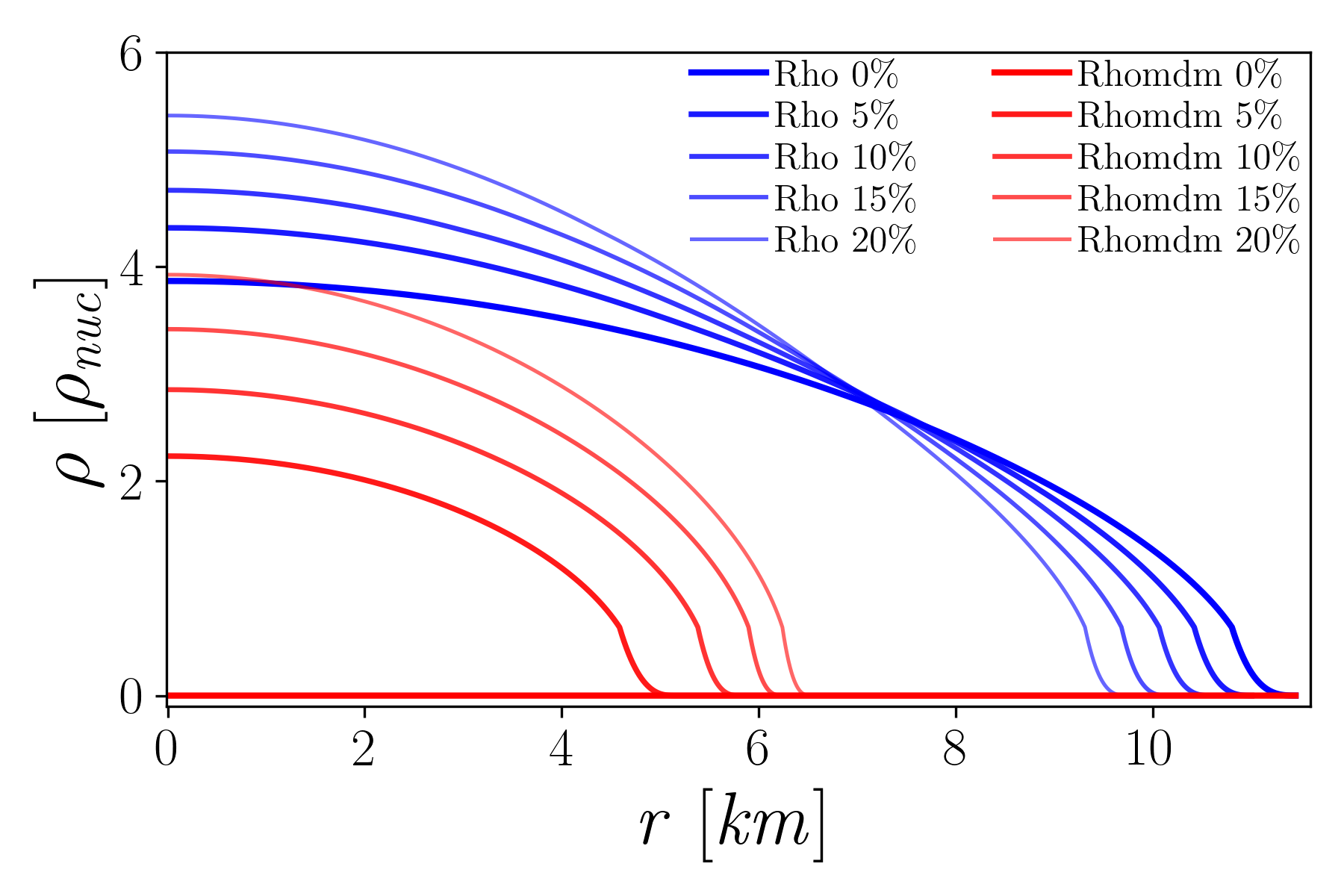}
\caption{Density profiles of admixed neutron stars with a 0\%, 5\%, 10\%, 15\% and 20\% mirror dark matter contribution to the total gravitational mass of 1.4 M$_\odot$. Blue lines refer to the baryonic density, while red ones denote the dark matter profiles. It is noticeable that with increasing dark matter content, the star has a more compact equilibrium configuration.
  \label{fig:density_dm}}
\end{center}
\end{figure} 

Once the integration has reached the boundary of the admixed star, the total gravitational mass $M$ is computed as the sum of the single masses, while the total-rest mass results from the integration of the summed up quantities 
\begin{linenomath}
\begin{equation}
    M_b= 4 \pi \int_{0}^{R} \frac{r^2\rho}{\sqrt{1-\frac{2M}{r}}}dr  ,
\end{equation}
\end{linenomath}
where the integral runs over the radial layers obtained in each integration step and $\rho=\textcolor{blue}{\rho_{\rm b}}+ \textcolor{red}{\rho_{\rm dm}}$. 

To validate our new algorithm, we test our TOV solver by comparing its results to the ones obtained with the methods of~\cite{Ciancarella:2020msu}. Figure \ref{fig:density_radius} shows an example of these comparisons, where mass-central density and mass-radius sequences are produced for the SLy EoS and overlayed, both for stars with 0\% and 50 \% dark matter content with respect to the total gravitational mass.  As shown in Figure \ref{fig:density_radius}, the results are equivalent for all the configurations and the comparisons are successful. As expected, the dark matter admixed neutron stars are less massive and more compact for equal central density values. The latter is shown in detail in Figure \ref{fig:density_dm}, where we compare the baryonic and dark matter density distributions, gradually increasing the dark matter contribution while keeping the total mass constant. The blue lines, used for the baryonic density, cross each other when increasing the dark matter percentage, allowing for the combination of higher central densities and smaller radii to reach hydrodynamical equilibrium. This results from the dark matter contributing only to the gravitational force, and not to the hydrodynamical pressure counteracting the gravitational contraction. 

\subsection{Single star test runs}

We perform convergence tests for various mirror dark matter percentages for different total gravitational masses keeping the mass constant for symmetric configurations, e.g., 25\% and 75\%, in order to test the dynamical stability of the implementation and the symmetric behaviour of the two components when switching their contribution, (Table \ref{tab:_TOV_convergency_test}). To do so, we run each initial configuration for four different resolution setups. These have respectively 64, 96, 128 and 160 grid points, with a related grid spacing of the coarsest refinement level of $2.0$, $1.\Bar{6}$, $1.0$ and $0.8$ M$_\odot$. We refer to these setups as R1, R2, R3, and R4. For all the setups a total of four refinement levels was used.
\begin{table}[htp!]
\begin{tabular}{|l|l|l|l|l|}
\hline
 Name              & Total mass ($M_{\odot}$) & Mirror dark matter \% & $\rho_c^b$   [\small{$\rho_{nuc}$}]            & $\rho_c^{dm}$   [\small{$\rho_{nuc}$}]      \\ \hline
SLy\_M127\_0pc  & 1.27                  & 0\%         & 3.558  & 0 \\ \hline
SLy\_M132\_25pc  & 1.32                  & 25\%         & 5.212      & 4.045      \\ \hline
SLy\_M14\_50pc & 1.40                  & 50\%        & 6.518      & 6.518      \\ \hline
SLy\_M132\_75pc  & 1.32                  & 75\%         & 4.045  &  5.212 \\ \hline
SLy\_M127\_100pc  & 1.27                  & 100\%         & 0  & 3.558 \\ \hline
\end{tabular}
\caption{Initial data, for convergence tests carried out with single mirror dark matter admixed neutron stars; cf. main text for further details.}
\label{tab:_TOV_convergency_test}
\end{table}

\begin{figure}[H]
\includegraphics[width=0.5\textwidth]{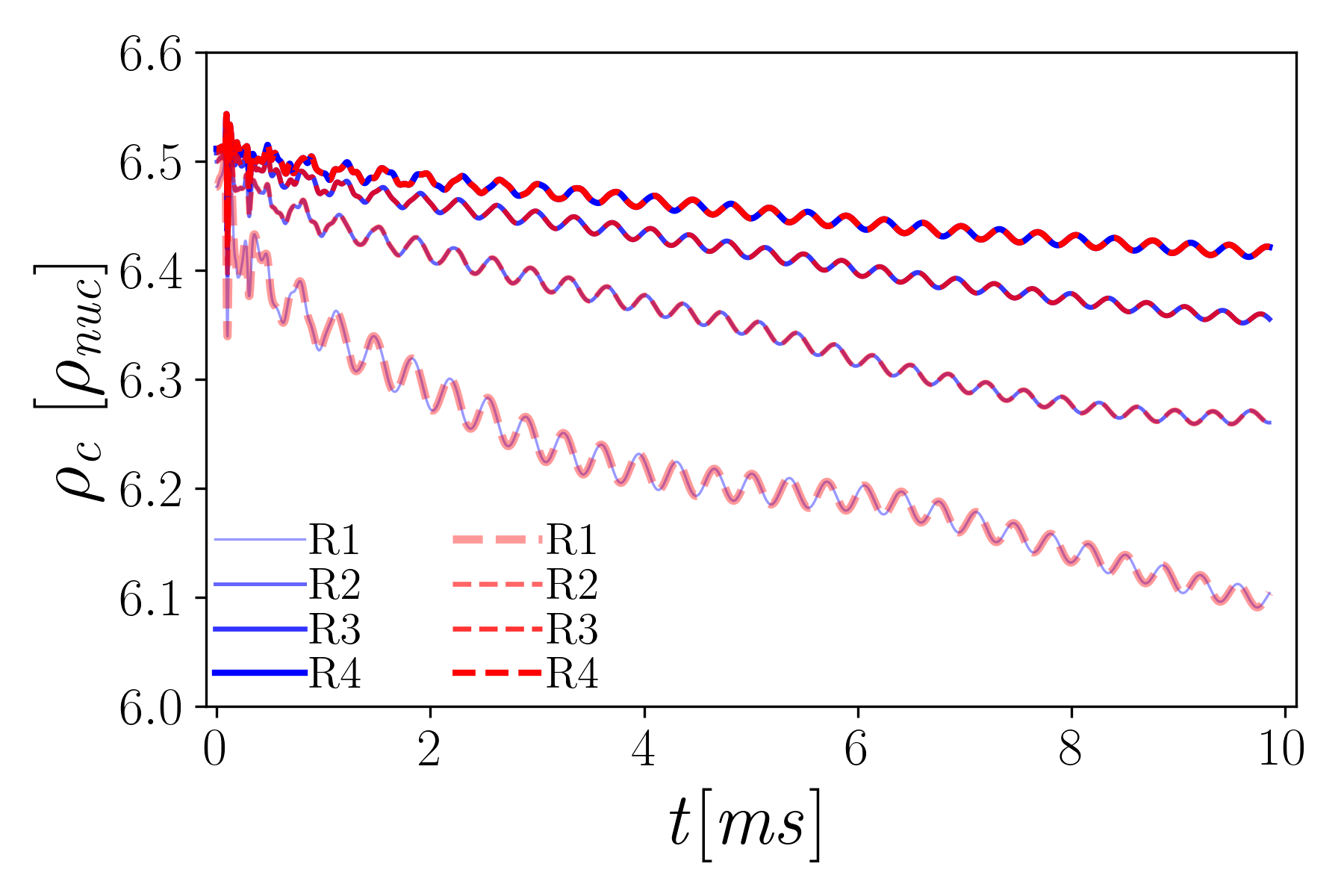}
\includegraphics[width=0.5\textwidth]{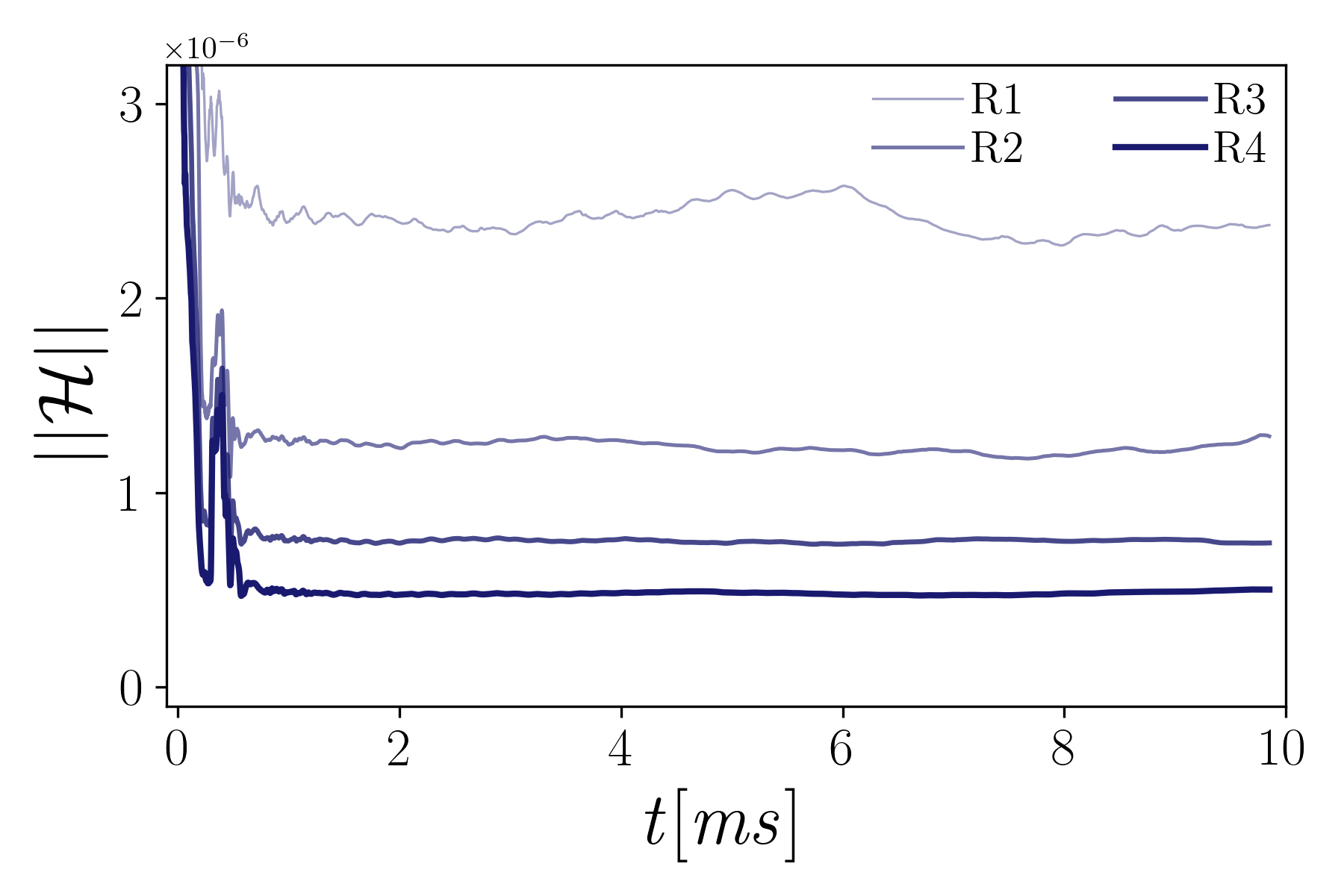}
\caption{\textbf{Left}: Time evolution of the central density of baryonic matter (blue) and mirror dark matter (red) for the four different resolutions presented in this section.
\textbf{Right}: Time evolution of the L2-norm of the Hamiltonian constraint for R1, R2, R3, and R4. The convergence order for the three highest resolutions is two. In all these runs the star is 50\% dark matter admixed and has a total gravitational mass of 1.4 M$_\odot$ and a baryonic mass of 1.56 M$_\odot$.} \label{fig:convergency_test}
\end{figure}

The results of one of these simulations, with a 50\% dark matter contribution and a 1.4 M$_\odot$ total gravitational mass, are shown in Figure~\ref{fig:convergency_test}. The evolution of the central density for baryonic and dark matter, on the left, is symmetric and their values show weaker oscillations with increasing resolutions. For the Hamiltonian constraint, in the right panels, we obtain a second order convergence using the three highest resolution runs. To test this, the convergence order of the central density for the same resolution was computed, which led to the same result. 

\section{Binary Neutron Star Simulations}
\label{sec:Bin}
In the following, we first describe how initial data for the binary neutron star simulations were constructed and present first simulation results. Specifically, we discuss the gravitational waveforms and the ejecta mass values.

\subsection{Initial configurations}

\begin{table}[H]
\begin{tabular}{|l|l|l|l|l|l|}
\hline
               & M$_{A,B}$ ($M_{\odot}$) & Mirror dark matter \% & $\rho_c^b$ [\small{$\rho_{nuc}$}]              & $\rho_c^{dm}$   [\small{$\rho_{nuc}$]}   & R$_{A,B}$ [\small{km}]   \\ \hline
SLy\_M14\_0  & 1.4                  & 0\%         & 3.866 & 0 & 11.45\\ \hline
SLy\_M14\_5  & 1.4                  & 5\%         & 4.360      & 2.234 & 11.00     \\ \hline
SLy\_M14\_10 & 1.4                  & 10\%        & 4.713      & 2.854  &  10.60  \\ \hline
SLy\_M13\_0  & 1.3                  & 0\%         & 3.624  & 0 & 11.46\\ \hline
SLy\_M13\_5  & 1.3                  & 5\%         & 4.058  & 2.087 &  11.04\\ \hline
SLy\_M13\_10 & 1.3                  & 10\%        & 4.366      & 2.679 &  10.63   \\ \hline
SLy\_M12\_0  & 1.2                  & 0\%         & 3.398      & 0 & 11.46\\ \hline
SLy\_M12\_5  & 1.2                  & 5\%         & 3.791  & 1.960 & 11.04\\ \hline
SLy\_M12\_10 & 1.2                  & 10\%        & 4.056      & 2.499 & 10.65   \\ \hline
\end{tabular}
\caption{Initial data for the mirror dark matter admixed binary neutron star simulations. The columns refer to the gravitational mass of each star, the dark matter percentage of the gravitational mass, the initial central density values for baryonic matter and dark matter respectively, and the total radius of the single stars. For the 1.4 M$_\odot$--1.4 M$_\odot$ and 5\%  configuration, runs were performed for the four different resolution setups described in Section 3.2. }
\label{binary_runs}
\end{table}

To enable the simulation of binary neutron stars, we use superimposed initial data. For this purpose, we solve the TOV equations separately for each star, apply a boost to both stars individually, and 'sum' the two separate spacetimes. Finally we also subtract the Minkowski spacetime, obtaining $g_{\mu \nu}= g_{\mu \nu}^{\text{star}_A}+g_{\mu \nu}^{\text{star}_B}-\eta_{\mu \nu}$. To ensure that the stars are on orbits with small eccentricity, we manually adjust the initial velocity using a shooting method.

This method allows to `construct' arbitrary configurations, but neither the Einstein Constraints nor the equations for hydrodynamical equilibrium are solved. Hence, spurious constraint violations and artificial density oscillation of the stars are present in simulations that use this kind of initial data. 
Nevertheless, this same approach was used to perform the first studies of eccentric binary neutron stars~\cite{Gold:2011df}, spinning binary neutron stars~\cite{Kastaun:2013mv}, and binary boson stars~\cite{Palenzuela_2007, Bezares:2022obu}. 
In the future, we aim to construct consistent initial data for dark matter admixed binary neutron stars. Given the complexity of the problem, we leave this to future work and use superimposed initial data for the time being.

\subsection{Time evolutions}

\begin{table}[htp!]
\centering
\begin{tabular}{|l|l|l|l|l|}
\hline
               & $n$ & $n_{mov}$  & $h^{max}$ (km)    & $h^{min}$(m) \\ \hline
R1   & 64               & 128        &   11.81  & 369 \\ \hline
R2   & 96               & 192        &   7.88   & 246 \\ \hline
R3   & 128              & 256        &   5.91   & 185 \\ \hline
R4   & 160              & 320        &   4.73   & 148 \\ \hline
\end{tabular}
\caption{Table of resolutions with columns showing the number of points per direction of the fixed levels and of the moving levels, the strain of the coarsest level and the strain of the finest one respectively. For all binary neutron star configurations we used 6 refinement levels.}
\label{tab:resolutions}
\end{table}

As a first step towards a qualitative investigation of binary neutron star mergers in the presence of mirror dark matter, we perform nine simulations of equal mass binaries. We take the three mass setups $1.2 M_\odot$--$1.2 M_\odot$, $1.3M_\odot$--$1.3M_\odot$, and $1.4M_\odot$--$1.4M_\odot$, and for each of these we consider stars with $0\%$, $5\%$, and $10\%$ dark matter contribution. All nine simulations are performed at the resolution R3; cf.\ Table~\ref{binary_runs}. In addition, we perform a resolution study for the $1.4M_\odot$--$1.4M_\odot$, $5\%$ dark matter system; cf.~Table~\ref{tab:resolutions}.  

\begin{figure}[htpb]
\begin{center}
\includegraphics[width=0.9 \textwidth]{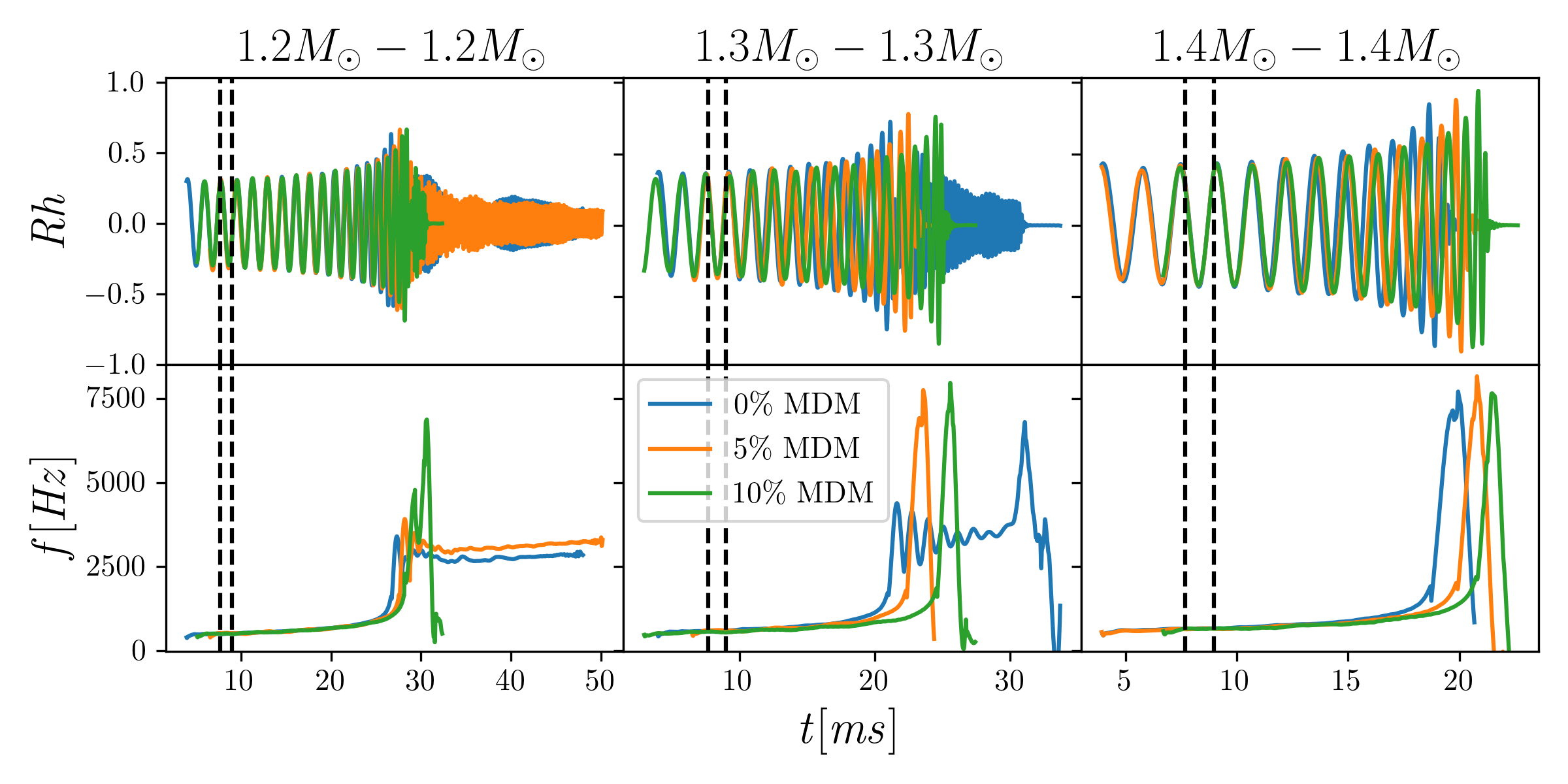}
\caption{Waveform strain and frequency of the $\ell=|m|=2$ GW mode for all simulations. Each plot contains simulations with the neutron star mass combination listed at the top of the column. The colors represent the different mirror dark matter fractions. All simulations with the same total mass are aligned during the first few orbits of the simulation time. Black dashed lines mark the time interval where the waveform alignment occurs.}
\label{gravitational}
\end{center}
\end{figure}

In Figure~\ref{gravitational}, we show the real part of the dominant mode of the GW signal for all the setups (top panels) and the evolution of the instantaneous GW frequency (bottom panels).  These panels encode the following behaviour of the binaries.
In the $1.4M_\odot$--$1.4M_\odot$ cases and in the dark matter admixed $1.3M_\odot$--$1.3M_\odot$ cases, the neutron stars collapse immediately into a black hole after merger.
In the 1.3 M$_\odot$--1.3 M$_\odot$ and 0\% dark matter case, a hypermassive~\cite{Baumgarte:2002jm} neutron star forms. This hypermassive neutron star survives for $\simeq$10 ms. The survival time reaches $\geq$20 ms for the $1.2 M_\odot$--$1.2 M_\odot$ case. For this `low-mass' scenario, also the 5\% and 10\% admixed stars form a long and short lived hypermassive neutron star respectively. 
From these results, we conclude that the addition of dark matter reduces the lifetime of the merger remnant and favors its prompt collapse into a black hole.

In addition, we find that for systems containing a higher percentage of dark matter the inspiral is `slower' and the merger frequency increases; cf. Table \ref{ejecta_table}.   We suggest that this is caused by the lower deformability of dark matter admixed neutron stars (all else being constant)~\cite{Ciancarella:2020msu}. 
However, further simulations using constraint-solved initial data are necessary to validate this observation. 

A key quantity in determining the brightness of the kilonova electromagnetic emission that may follow a binary neutron star merger is the mass of the ejected material. In Table~\ref{ejecta_table}, we list the properties of the ejected matter for all our simulations. Following the ballistic criterion, fluid elements are marked as unbound when $u_t<-1$ and $v_r>0$, where $u_t$ and $v_r$ are respectively the time component of the four-velocity and the radial velocity, i.e., we assume that fluid elements follow geodesics and are outward moving. 

\begin{table}[htpb]
\begin{tabular}{|l|l|l|l|l|}
\hline
 & $M_{ej}$ sphere ($M_{\odot}$) & $M_{ej}$ integral ($M_{\odot}$) & $M_{disk}$ ($M_{\odot}$) & $f_{merger} [Hz]$ \\ \hline

SLy\_M14\_0         &   -           &   -         &  0.001            & 1770  \\ 
SLy\_M14\_5         &   -           &   -         &  0.0008           & 2030  \\ 
SLy\_M14\_10        &   -           &   -         &  0.0014           & 2058  \\ \hline
SLy\_M13\_0         &   0.0168      &   4.8 $\cdot$ 10$^{-3}$    &  0.062            & 1817  \\ 
SLy\_M13\_5         &   0           &   0.7 $\cdot$ 10$^{-3}$   &  0.001            & 1910  \\
SLy\_M13\_10        &   0           &   0.8 $\cdot$ 10$^{-3}$   &  0.0006           & 2221  \\ \hline
SLy\_M12\_0         &   0           &   0.3 $\cdot$ 10$^{-3}$   &  $0.19\text{*}$   & 1746  \\ 
SLy\_M12\_5         &   0.0016      &   2.6 $\cdot$ 10$^{-3}$    &  $0.16\text{*}$   & 1818  \\ 
SLy\_M12\_10        &   0.0027      &   3.3 $\cdot$ 10$^{-3}$    &  0.017            & 2198  \\ \hline
\end{tabular} 
\caption{\textit{First column:} Name of the simulation. \textit{Second column:} Total unbound matter flowed through a sphere with radius 450 km centered around the origin. Unbound matter is identified using the ballistic criterion, (see main text).
\textit{Third column:} Maximum value of the integral of unbound matter over the simulation domain reached during the simulation time. Unbound matter is again identified using the balistic criterion.  \textit{Fourth column:} Mass of the disk at 5 ms after merger identified as the total bound mass outside a radius of 12 km from the center. \textit{Fifth column:} Frequency of the $\ell=|m|=2$ mode of the gravitational waveform taken at the merger, i.e., the time of the GW strain maximum. \\
\text{*} The disk of these simulations has been measured before the collapse of the remnant. Therefore it can be affected by a higher uncertainty due to the ambiguity in the definition of the disk.}
\label{ejecta_table}
\end{table}

\begin{figure}[htp!]
\begin{center}
\includegraphics[width=0.497 \textwidth]{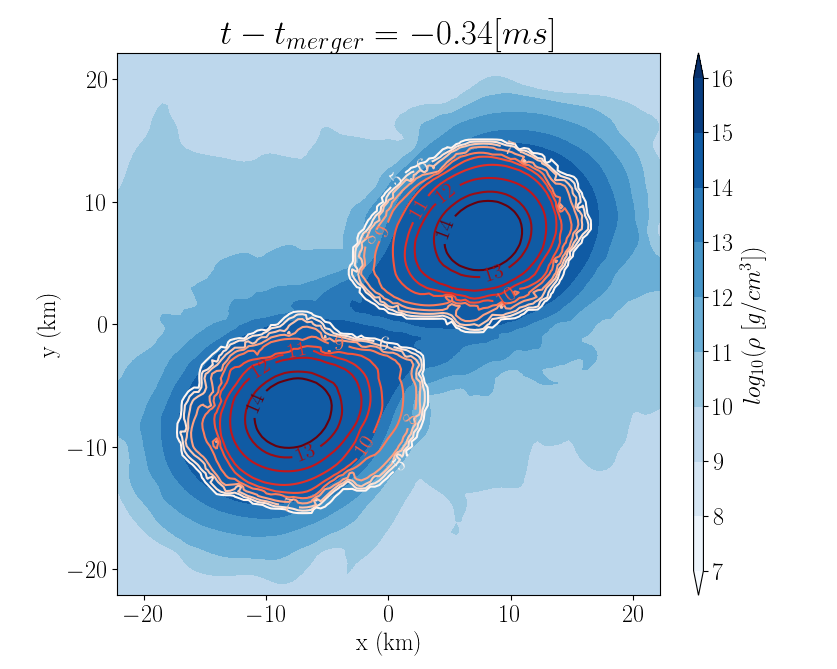}
\includegraphics[width=0.497 \textwidth]{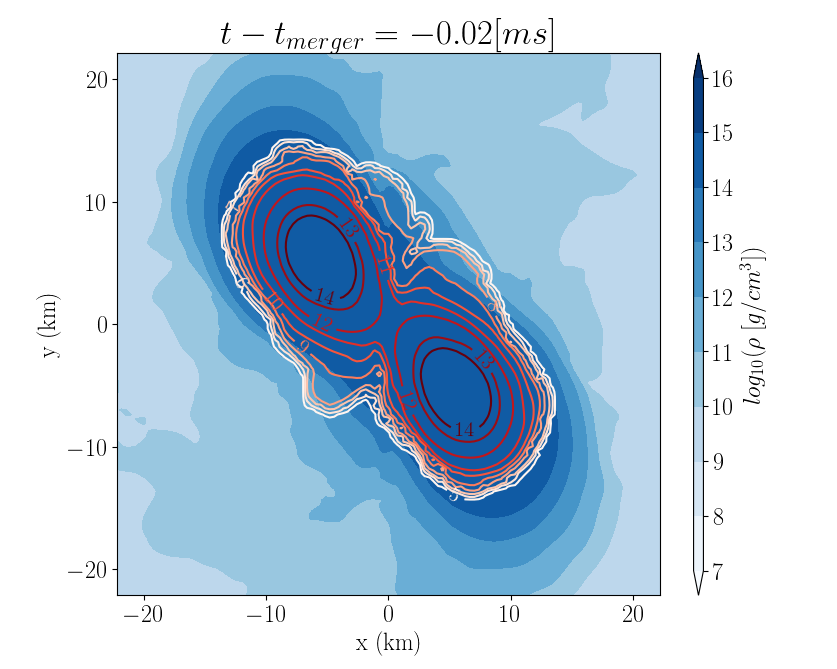}
\includegraphics[width=0.497 \textwidth]{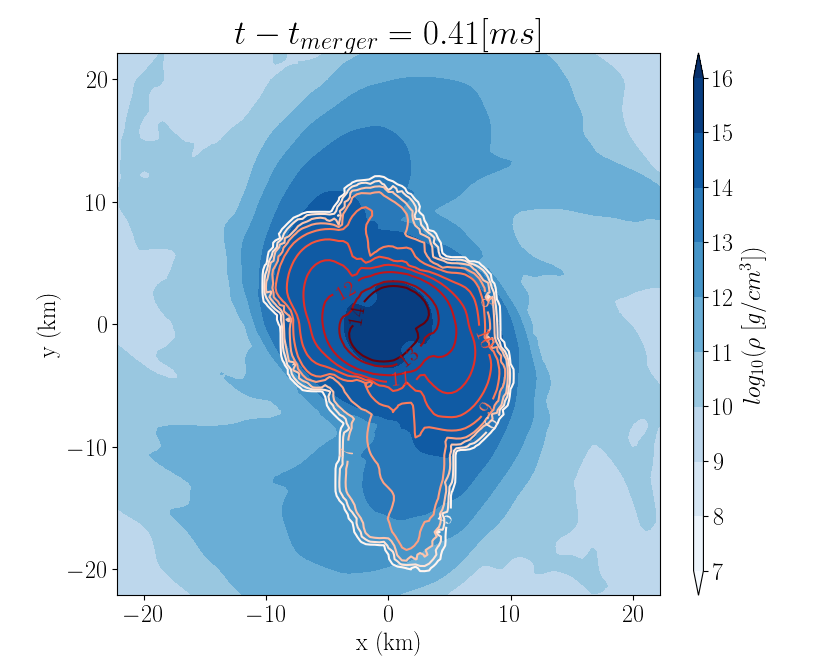}
\includegraphics[width=0.497 \textwidth]{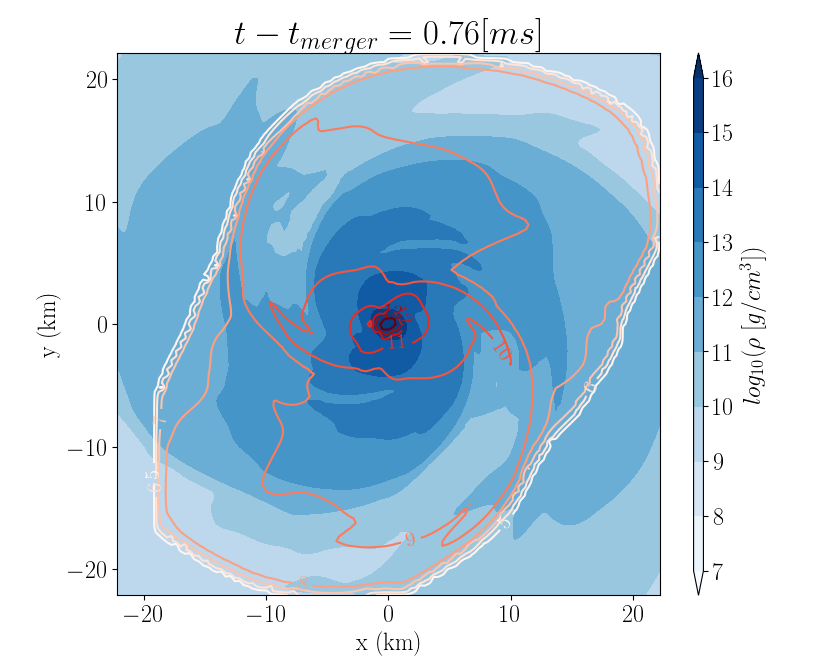}
\includegraphics[width=0.497 \textwidth]{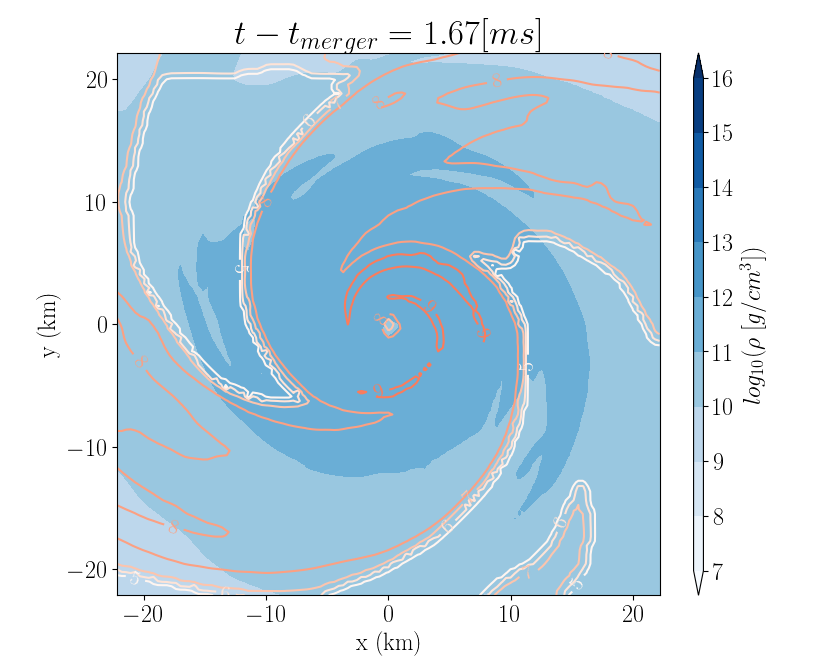}
\includegraphics[width=0.497 \textwidth]{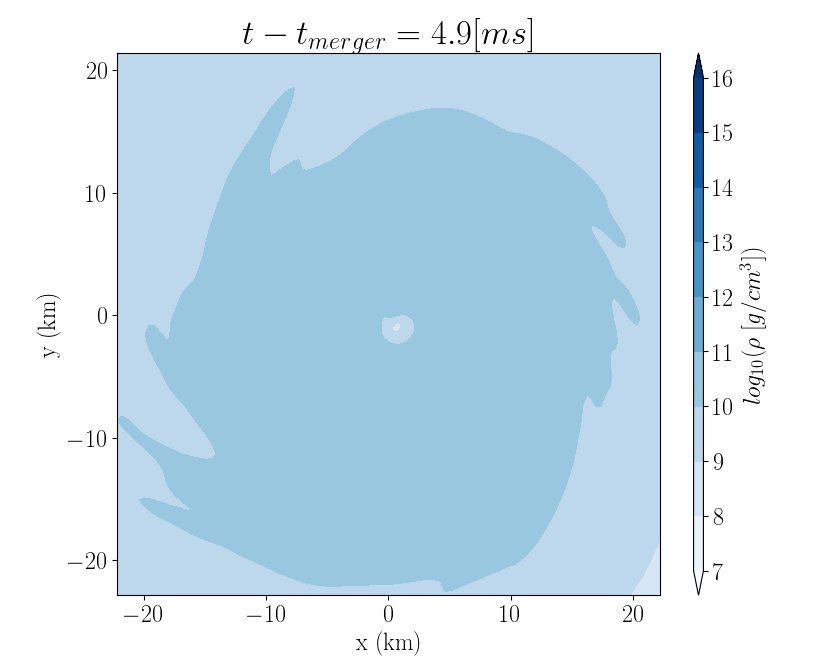}
\caption{Snapshots on the equatorial plane of the 1.3 M$_{\odot}$ -- 1.3 M$_{\odot}$  simulation with $5\%$ dark matter. $x$ and $y$ are in isotropic coordinates.  The colormap represents the $\log_{10}$ of baryonic rest mass density in cgs units. The contours represent the same quantity for mirror dark matter. }
\label{merger_evolution}
\end{center}
\end{figure} 

We use two different measures to extract the ejecta mass~\cite{Chaurasia:2018zhg}. In the first approach, we compute the mass flow of unbound matter through a sphere of coordinate radius $450$ km. The second approach uses a volume integration over the entire simulation domain. While the latter approach allows us to provide ejecta mass estimates even close to the merger remnant, the volume integral is sensitive to the artificial atmosphere, i.e., outward moving matter that decompressed will potentially be set to atmosphere and hence not tracked properly. 
On the other hand, when computing the ejecta mass with respect to the matter flow through spheres, one would need longer simulation times to ensure that also the slow moving ejecta reaches the extraction radius of 450 km during the simulation time. Given these pros and cons, we determine the qualitative behavior of the ejecta mass using both methods, as reported in Tab.~\ref{ejecta_table}\footnote{We  emphasize once more that, for a full quantitative analysis, we would need to use constraint solved initial data rather than simplified superimposed initial configurations.}. Unfortunately, we can not find a clear pattern for the ejecta mass for all our simulations. While it seems that an increasing dark matter fraction decreases the ejecta mass for the $1.3M_\odot$--$1.3M_\odot$ setup, the opposite holds for the $1.2M_\odot$--$1.2M_\odot$ setup.

In addition to the investigation of the ejecta mass, we also compute the disk mass of the systems (Tab.~\ref{ejecta_table}). The disk mass value is computed $5$ $\rm ms$ after the formation of the black hole, or at the end of the simulation when no black hole is formed. In the latter scenario, the disk is defined as the total bound mass located outside a sphere of radius 12 km (with its center at the coordinate origin). Overall, we find that the mass of the disk decreases as we increase the dark matter percentage, except for the $1.4M_\odot$--$1.4M_\odot$ case. We suggest that this is caused by the fact that as the content of mirror dark matter increases, the neutron stars become more compact and it becomes harder to separate material from the bulk of the stars. Given that the disk mass is a key quantity for determining possible disk wind ejecta~\cite{Siegel:2017nub,Fernandez:2012kh} and due to the overall small dynamical ejecta, we conclude that mirror dark matter admixed binary neutron star mergers will generally lead to dimmer electromagnetic signatures than `pure' baryonic matter binary neutron star mergers.

In Figure \ref{merger_evolution}, we show snapshots of the different phases of the coalescence taken from the simulation with $1.3 M_\odot$--$1.3 M_\odot$ neutron stars and 5\% mirror dark matter. The colors indicate the density of baryonic matter in log$_{10}$ scale, while the contours define the same quantity for the dark matter component. We see that mirror dark matter remains bound in the stars core even during the last phases of the inspiral with little or no tidal ejecta. Shortly after merger we observe a shock wave that ejects some dark matter into the disk. However,
such matter is strongly bound to the remnant and rapidly falls back into itself in a short time leaving no dark matter outside the black hole at the end of the simulation.

\section{Conclusions}
\label{sec:Conc}

We upgraded the numerical-relativity code BAM to allow for the simulation of neutron stars admixed with mirror dark matter. To achieve this, we implemented a two fluid approach in which the two fluids only interact via gravity. For this purpose, we changed BAM's TOV solver and the routines that handle general-relativistic hydrodynamics. We validated our new routines by comparing our simulations with results obtained as described by Ciancarella et al.~\cite{Ciancarella:2020msu} and found perfect agreement. We also performed a convergence study for the single star case and proved the stability of our implementation. 

As a first application of our new code version, we performed binary simulations of mirror dark matter admixed neutron stars and compared them against simulations of binary neutron star mergers in the absence of dark matter. Although we used simplified initial data by superimposing the spacetimes of two distinct TOV solutions, our results show that increasing the amount of dark matter present in the neutron stars translates into a longer inspiral (keeping all other system parameters unchanged), likely due to a lower deformability of dark matter admixed neutron stars.  

In addition, we also computed the ejecta mass and the disk mass for our simulations. While we obtained no clear pattern for the ejecta mass, we noticed a decrease of the disk mass for an increasing dark matter percentage inside the stars. We suspect this to be connected to the faster formation of the black hole after the merger and that it becomes progressively harder to eject material from the bulk of the stars prior to the black hole formation. Similarly, the lack of dark matter ejecta and debris disks is related to its concentration in the core of the neutron star and its low percentage for the cases that we investigated.

While our work is an important step towards modelling the gravitational-wave and electromagnetic emission of neutron star binaries admixed with dark matter, we also found shortcomings of our first simulations. Most notably, the superimposed initial data affects the simulations such that no clear convergence can be observed. To overcome this issue, we plan to continue our studies by improving our determination of the initial data with a proper description of binary dark matter admixed neutron stars in hydrodynamical equilibrium and in agreement with Einstein Equations. This will enable a quantitative investigation of this interesting class of physical objects. 

\appendix

\section*{Appendix: Resolution study for the binary neutron star simulations}

\begin{figure}[htp!]
\begin{center}
\includegraphics[width=0.48 \textwidth]{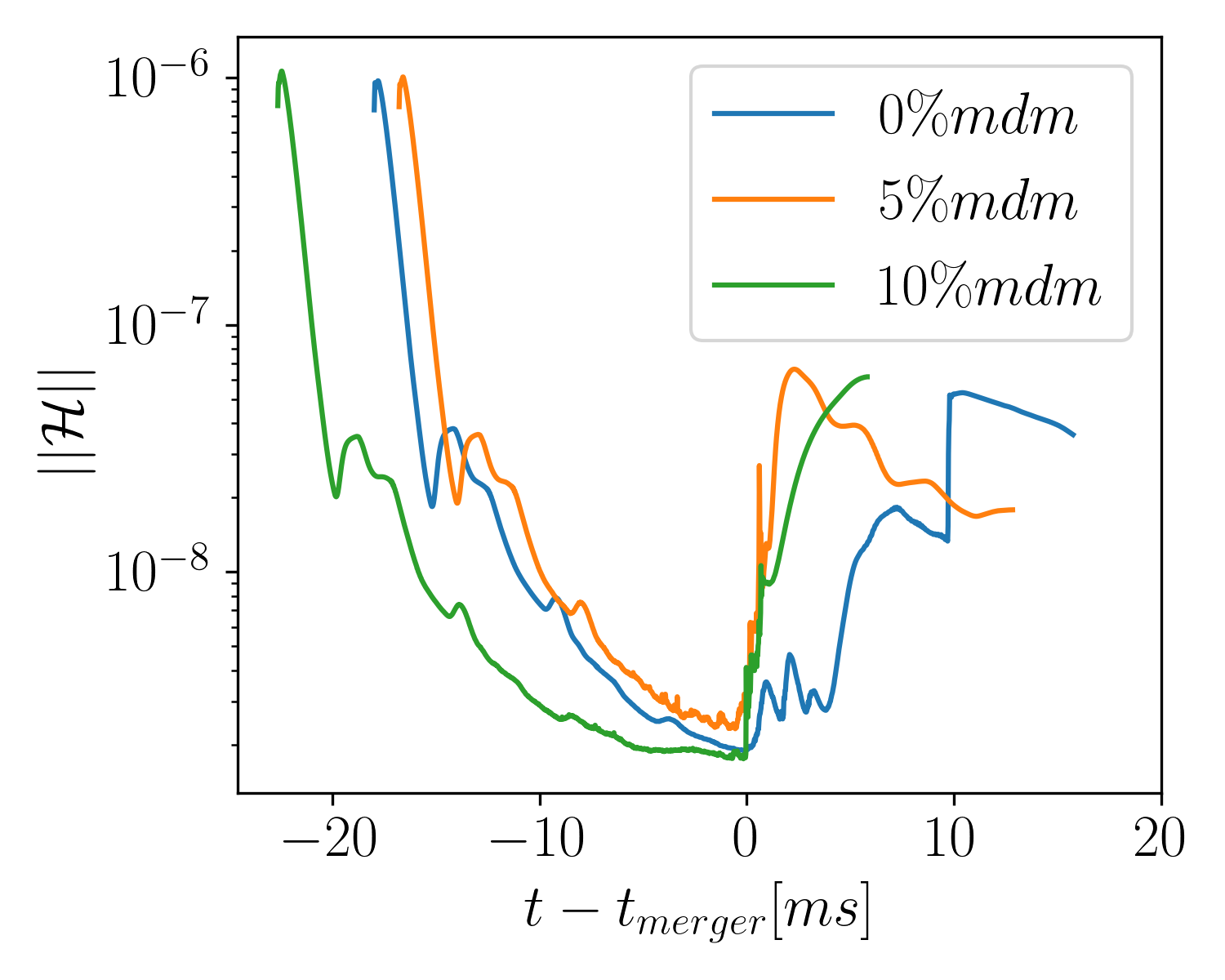}
\includegraphics[width=0.48 \textwidth]{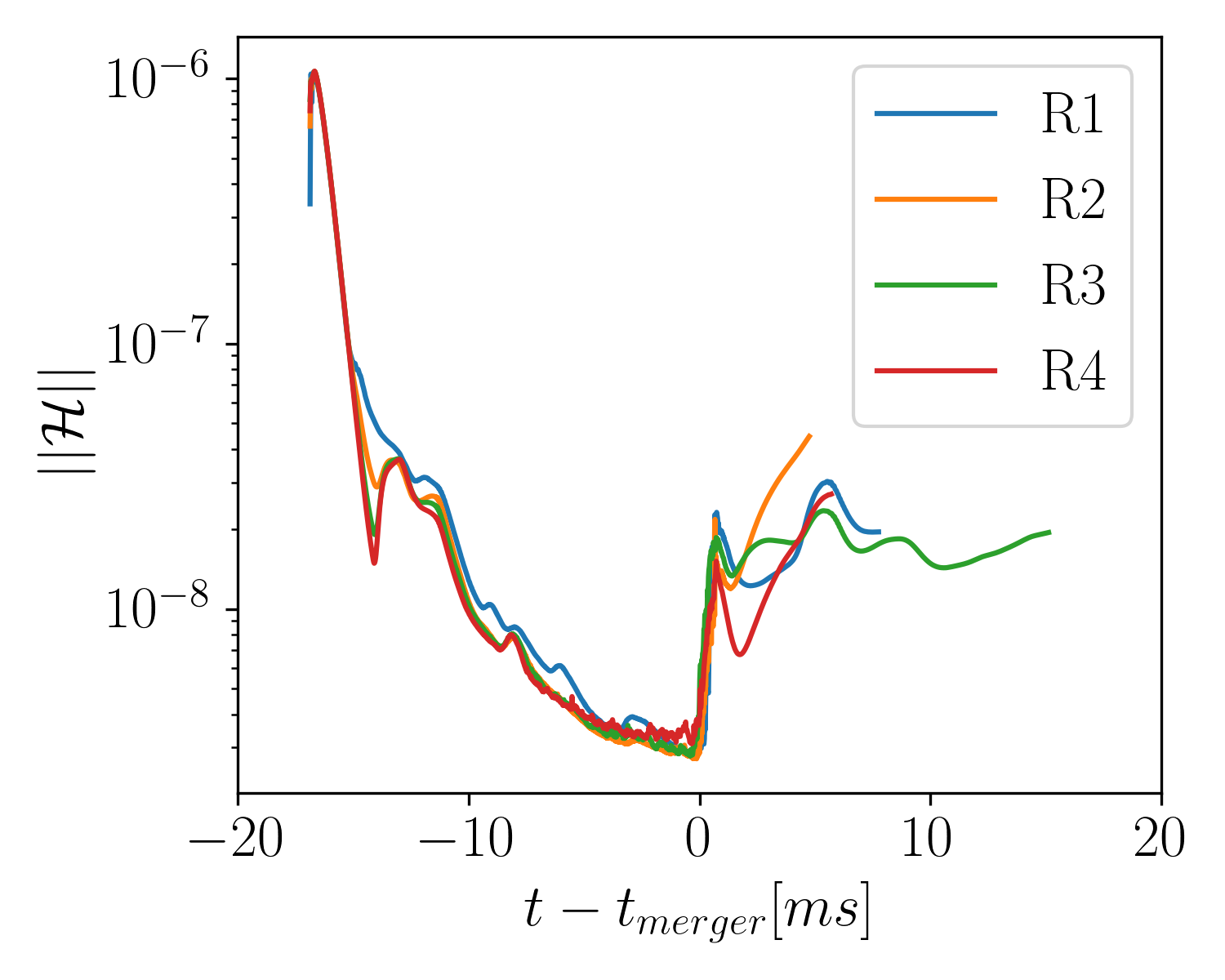}
\caption{\textbf{Left}: Evolution of Hamiltonian norm for the $1.3$ M$_{\odot}$--$1.3$ M$_{\odot}$ simulations on the coarsest level. Overall, we find a decreasing constraint violation before the merger, due to the constraint damping properties of the Z4c evolution scheme. The jumps at late times are due to the merger and then the formation of a singularity. \textbf{Right}: Hamiltonian norm on the coarsest level for the simulation of $1.4M_{\odot}$--$1.4M_{\odot}$ and $5\%$ mirror dark matter stars for the four different resolutions described in Table \ref{tab:resolutions}.}
\label{ham_resolution}
\end{center}
\end{figure}

In Figure~\ref{ham_resolution}, we show the Hamiltonian constraint for the $1.3 M_\odot$-$1.3 M_\odot$ binaries (left panel) and the Hamiltonian constraint for the four different resolutions used in the $1.4 M_\odot$-$1.4 M_\odot$ study for the 5\% mirror dark matter configuration (right panel). Due to the simplified approach adopted herein to produce initial data, the Hamiltonian constraint violation is initially high and is then damped as an effect of the use of the Z4c formulation of the ADM equations. We find that this initial constraint violation dominates the entire simulation. Therefore, increasing the resolution during the evolution leads to no meaningful decrease of the Hamiltonian constraint (right panel). The jumps in the left plot correspond to the collapse into a black hole, which for the dark matter admixed cases forms immediately after the merger. In the 0\% setup the jump is at 10 ms after the merger, confirming the formation of a hypermassive neutron star before the final collapse, as seen in Figure~\ref{gravitational}. 

\vspace{6pt} 

\authorcontributions{M.E. and F.S. carried out simulations, data analysis and  preparation of figures. F.P. and T.D. proposed the idea and contributed to the code validation and data interpretation. V.S. performed data verification. All authors have contributed to the preparation of the manuscript and discussions.}

\funding{V.S. acknowledges the support from the FCT (Funda\c c\~ao para a Ci\^encia e Tecnologia I.P, Portugal) under the projects No. EXPL/FIS-AST/0735/2021, No. UID/FIS/FIS/04564/2020, No. UIDP/04564/2020 and No. UIDB/04564/2020.
The simulations were performed on the national supercomputer HPE Apollo Hawk at the High Performance Computing (HPC) Center Stuttgart (HLRS) under the grant number GWanalysis/44189, 
on the GCS Supercomputer SuperMUC at Leibniz Supercomputing Centre (LRZ) [project pn29ba], the HPC systems Lise/Emmy of the North German Supercomputing Alliance (HLRN) [project bbp00049], and the Minerva cluster of the Max Planck Institute for Gravitational Physics.}

\acknowledgments{We thank the `extreme matter' subgroup of the LIGO-Virgo-KAGRA Collaboration for valuable feedback. 
}

\conflictsofinterest{The authors declare no conflict of interest.}

\begin{adjustwidth}{-\extralength}{0cm}

\reftitle{References}

\bibliography{paper.bbl}
\end{adjustwidth}
\end{document}